\documentclass[]{aa}  

\usepackage{graphicx}
\usepackage{txfonts}
\usepackage{xcolor}
\usepackage{hyperref}
\makeatletter
\renewcommand*\aa@pageof{, page \thepage{} of \pageref*{LastPage}}
\makeatother
\newcommand{\teff}{T$_{\rm eff}$ }
\newcommand{\logg}{$\log(g)$ }
\newcommand{\feh}{[Fe/H] }
\newcommand{\vmic}{$\xi_{\rm mic}$ }

\newcommand{\fstar}{Z06-star}
\newcommand{\jstar}{J13-star}

\begin{document} 
\title{The first r-process enhanced star\\ confirmed to be a member of the Galactic bulge}


   \author{R. Forsberg
          \inst{1}
          \and
          R. M. Rich\inst{2,3}
          \and
          N. Nieuwmunster\inst{1,3} 
          \and
          H. Jönsson\inst{4}
          \and
          M. Schultheis \inst{3} 
          \and
          N. Ryde\inst{1} 
          \and 
          B. Thorsbro\inst{1,5} 
          }

   \institute{Lund Observatory, Department of Astronomy and Theoretical Physics, Lund University, Box 43, SE-22100 Lund, Sweden\\
              \email{rebecca@astro.lu.se}
    \and         
    Department of Physics and Astronomy, ICLA, 430 Portola Plaza, Box 951547, Los Angeles, CA 90095-1547 
    \and 
    Université Côte d’Azur, Observatoire de la Côte d’Azur, Laboratoire Lagrange, CNRS, Blvd de l’Observatoire, 06304 Nice, France 
    \and 
    Materials Science and Applied Mathematics, Malm\"o University, SE-205 06 Malm\"o, Sweden 
    \and
    Department of Astronomy, School of Science, The University of Tokyo, 7-3-1 Hongo, Bunkyo-ku, Tokyo 113-0033, Japan 
    }

   \date{Received June 20, 2022; accepted October 10, 2022}

 
  \abstract
   {}
   {Stars with strong enhancements of r-process elements are rare and tend to be metal-poor, with generally [Fe/H] $<-2$\,dex and found in the halo. In this work we aim to investigate a candidate r-process enriched bulge star with a relatively high metallicity of [Fe/H] $\sim -0.65$\,dex, and compare it with a previously published r-rich candidate star in the bulge.}
   {We reconsider the abundance analysis of a high-resolution optical spectrum of the red-giant star 2MASS J18082459-2548444 and determine its europium (Eu) and molybdenum (Mo) abundance, using stellar parameters from five different previous studies. Applying 2MASS photometry, Gaia astrometry and kinematics, we estimate distance, orbits, and population membership of 2MASS J18082459-2548444 and a previously reported r-enriched star 2MASS J18174532–3353235.}
   {We find that 2MASS J18082459-2548444 is a relatively metal rich enriched r-process star that is enhanced in Eu and Mo but not substantially enhanced in s-process elements. It has a high probability of membership in the Galactic bulge based on its distance and orbit. We find that both stars show r-process enhancement with elevated [Eu/Fe]-values, even though 2MASS J18174532–3353235 is 1\,dex lower in metallicity. Additionally, we find that 2MASS J18174532–3353235 plausibly has a halo or thick disc origin.}
   {We conclude that 2MASS J18082459-2548444 represents the first example of a confirmed r-process enhanced star confined to the inner bulge, possibly a relic from a period of enrichment associated with the formation of the bar.}


   \maketitle

%
\section{Introduction}
The composition of stars present a record of the formation and chemical enrichment of the Milky Way. Stars form from molecular clouds, and potentially carry a chemical fingerprint of their birth cloud that may include pollution from earlier events. The Galactic interstellar medium is chemically enriched over time, through nucleosynthetic processes, occurring either in stars or in explosive environment such as supernovae (SNe) and neutron star mergers. 

The Galactic bulge is of great interest because of its age and high metallicity, consistent with early, rapid, enrichment. The modern era of study gave observations of high iron abundance and $\alpha$-enhancements \citep{rich88,mcwilliamrich1994ApJS...91..749M} paired with old age \citep{ortolani95} substantiated this picture of rapid enrichment via Type II SNe \citep{matteucci1990}. The actual picture may be more nuanced. Deep studies of the main sequence in fields imaged by the Hubble Space Telescope largely support a predominant older age \citep{clarkson08,renzini2018} while \citet{bensby17} argues for a far greater dispersion in age, based on the analysis of microlensed bulge dwarfs. The recent study of \cite{joyce22} confirms that the most metal rich stars may be younger than 10 Gyr, with a 5 Gyr age dispersion. 

Stars with [Fe/H]\,$>-0.5$\,dex are predominantly in a bar \citep{ness2012ApJ...756...22N,ness13} that appears to have formed via the buckling of a pre-existing disc \citep{shen10,dimatteo2014A&A...567A.122D}. Overall, the present day finds a tension between apparently robust evidence for early rapid enrichment from analysis of the color-magnitude diagrams and luminosity functions, with the study of the microlensed dwarfs providing a counterpoint of possibly extended formation. The progressive flattening of the metal rich population \citep{johnson2020} offers some support for the extended formation scenario.  


This tension between evidence of early enrichment in the bulge versus a potentially extended formation history for the most metal rich bulge populations is interesting. One characteristic of the ancient Galactic halo is the presence of neutron-capture \citep{burbidge1957} enhanced stars, especially those with high amount of r-process elements. The discovery of a bulge/inner halo candidate displaying the full r-process pattern \citep{johnson2013ApJ...775L..27J} raises the possibility that such stars also formed early in the history of the bulge or inner halo. 

These types of r-enhanced stars remain rare, but are of sufficient interest that they are catalogued and studied across Galactic populations with more than 300 known at this time \citep[see the R-process Alliance,][]{holmbeck2020ApJS..249...30H}. Whilst the origin of such enhancements remains an open matter of discussion to date, and it is possible that the primary source has changed over time \citep{sneden2008ARA&A..46..241S}, they are proposed to offer clear signatures of r-process events. Such events could be neutron star mergers \citep[NSM,][]{matteucci2014MNRAS.438.2177M} and/or magneto-rotational driven SNe \citep[MRSNe,][]{nishimura2006ApJ...642..410N,kobayashi2020ApJ...900..179K} which can provide the neutron-rich conditions that favour the nucleogenesis of these heavy elements. 

In general, stars showing r-process enhancements are metal poor, [Fe/H]\,$<-2$\,dex \citep{mcwilliam1995AJ....109.2757M,sneden2003ApJ...591..936S,sneden2008ARA&A..46..241S,holmbeck2020ApJS..249...30H}, appearing in the Milky Way halo and in dwarf galaxies \citep{hirai2015,matsuno2021A&A...650A.110M,jeon2021MNRAS.506.1850J}. These are classified into two groups, r-I and r-II, depending on the level of r-process enrichment. The discrepancy of r-I and r-II stars has been given by $0.3<$\,[Eu/Fe]\,$\leq\,1.0$\,dex and [Eu/Fe]\,$>1.0$\,dex, respectively, with [Ba/Eu]\,$<$\,0\,dex for the r-II stars \citep[][]{Christlieb2004A&A...428.1027C}. In the R-process Alliance they list 232 r-I and 72 r-II stars, and propose a new demarcation between these two groups at [Eu/Fe]\,=\,0.7\,dex  \citep{holmbeck2020ApJS..249...30H}. As the astrophysical site(s) of the r-process remains a matter of vigorous debate \citep[see e.g.][]{cote2019ApJ...875..106C}, it continues to be important to document such stars in a range of stellar populations as their presence potentially gives insight into the early enrichment history of a population.


One interesting set of spectra was first used in \citet{zoccali2006} and later reanalysed in several subsequent studies \citep{lecureur2007,ryde2010,vanderswaelmen,jonsson2017b,lomaeva2019,Forsberg2019,forsbergsubmitted}. A star found in the set, K-giant 2MASS J18082459-2548444 (hereafter we call it the \fstar), has recently been shown to be high in [Eu/Fe] \citep{vanderswaelmen,Forsberg2019} and [Mo/Fe] \citep{forsbergsubmitted}. At [Fe/H]\,$\sim\,-0.65$\,dex \citep{lecureur2007,zoccali2008,ryde2010, johnson2014AJ....148...67J,jonsson2017b}, this star is remarkably high in metallicity for a potentially r-process enhanced star. While at present we report on only this star, the discovery of it makes emphatic the importance of searching for other stars with unusual enhancements of heavy elements, also in the Galactic bulge. 

Indeed, \citet{johnson2013ApJ...775L..27J} publish the first r-enriched candidate in the Galactic bulge, 2MASS J18174532–3353235, located at $(l,b)=(-1, -8)^{\circ}$. The star is reported to have a relatively high metallicity (relative the halo r-enhanced population) of [Fe/H]\,=\,$-1.67$\,dex and an r-enrichment as traced by europium of [Eu/Fe]\,=\,1.0\,dex and molybdenum of [Mo/Fe]\,=\,0.72\,dex. It should be noted that \citet{johnson2013ApJ...775L..27J} do not rule out a halo origin of 2MASS J18174532–3353235 (hereafter we call it the \jstar) and note that with higher precision proper motions, the kinematics of the star would be constrained with greater certainty. 

In this work we re-analyse the optical spectrum of the \fstar, as described in Sect.~\ref{sec: data}, using the stellar parameters from the earlier studies \citep{zoccali2006,lecureur2007,ryde2010,vanderswaelmen,jonsson2017b} to determine the abundances of the r-process elements europium and molybdenum, as described in Sect.~\ref{sec: analysis if spectrum}. We compare the results with the studies by \citet{Forsberg2019,forsbergsubmitted}.

Further, in Sect.~\ref{sec: bulge membership investigation}, we investigate the bulge membership of the two r-enhanced stars, the \fstar, 2MASS J18082459-2548444, and the \jstar, 2MASS J18174532–3353235. With the data from Gaia early-data release 3 \citep[Gaia EDR3,][]{gaiacollaboration2021A&A...649A...1G} we now have a high precision value for the proper motion of these stars, which allows us to calculate their orbits.

\section{Observational data}
\label{sec: data}
\begin{figure}[h]
    \centering 
    \includegraphics[width=\hsize]{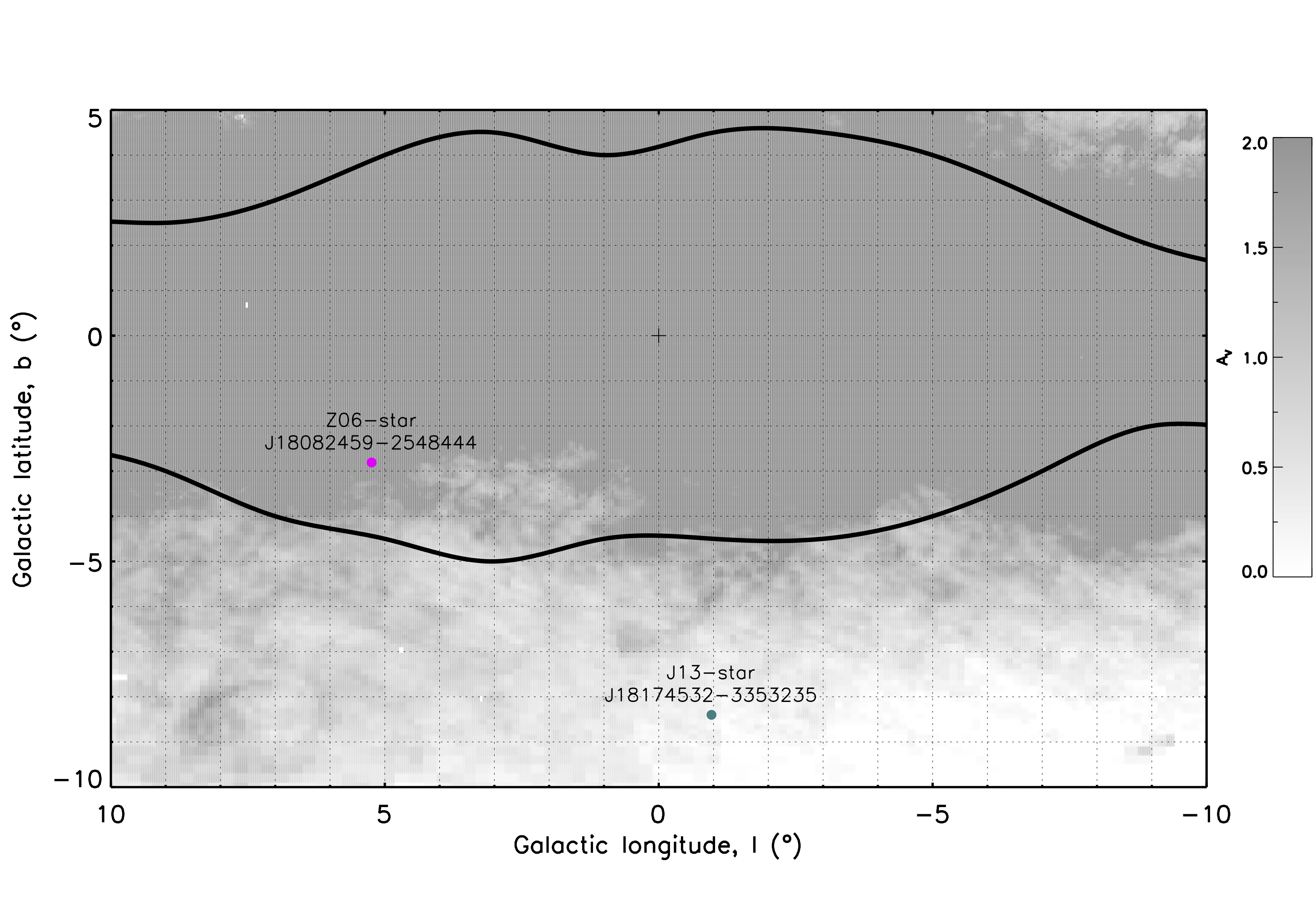}
    \caption{Map of the Galactic bulge showing the position of the \fstar, 2MASS J18082459-2548444 (magenta) and the \jstar, 2MASS J18174532–3353235 (teal), presented in this work. The dust extinction towards the bulge is taken from \citet{gonzalez:11,gonzalez2012} and scaled to optical extinction \citep{cardelli}. The scale saturates at A$_{\text{V}} = 2$, setting the upper limit. We also plot the COBE/DIRBE outlines of the Galactic bulge \citep{weiland:94}.}
    \label{fig:bulgefields}
\end{figure}

The spectrum of the \fstar\ is part of the \citet{zoccali2006} data set and has been observed with the FLAMES-UVES (R $\sim 47,000$) at the European Southern Observatory's' Very Large Telescope (ESO's VLT). The spectral wavelength is limited to 5800-6800\,Å. The star is located at (l,b) = ($5.2^{\circ}$,$-2.8^{\circ}$) placing it in the line of sight within the bulge as outlined from COBE/DIRBE \citep{weiland:94}, as seen in Fig.~\ref{fig:bulgefields}. 

The \fstar\ has stellar parameters determined in \citet{lecureur2007,zoccali2008,ryde2010,johnson2014AJ....148...67J,jonsson2017b}, see Table~\ref{table: stellar parameters}. With an effective temperature of $\sim$4400\,K and $\log(g)$ of $\sim$1.75, it is a typical red-giant star. With the combined works of \citet{lecureur2007,ryde2010,johnson2014AJ....148...67J,vanderswaelmen,jonsson2017b,lomaeva2019,Forsberg2019,forsbergsubmitted}, the known chemical fingerprint of the \fstar\ consists of 21 abundances, see Table \ref{table:norm abundances}.

The \jstar, placed at (l,b) = ($1^{\circ}$,$-8^{\circ})$ is also marked in Fig.~\ref{fig:bulgefields} and is found within the bulge Plaut field. The spectrum of the \jstar, which has been analysed in \citet{johnson2013ApJ...775L..27J}, has been observed with the Magellan Inamori Kyocera Echelle (MIKE) spectrograph on the 6.5\,m Magellan Clay telescope (R $\sim 30,000$).

\section{Analysis of spectrum}
\label{sec: analysis if spectrum}
In this section, we give a brief discussion of the analysis of the spectrum of the \fstar, where we follow the methodology as outlined in the Jönsson-series \citep{jonssona2017A&A...598A.100J,jonsson2017b,lomaeva2019,Forsberg2019,forsbergsubmitted}. We re-determine the Mo and Eu abundances using the stellar parameters presented in \citet{lecureur2007,zoccali2008,ryde2010,johnson2014AJ....148...67J} to confirm the high abundances determined in previous papers. Readers are referred to these papers for a more detailed description of the stellar parameters and details of the elemental abundance determination for the other elements presented in Table \ref{table:norm abundances}. 

The analysis of the spectra is done using the tool Spectroscopy Made Easy \citep[SME, version 554][]{valentisme1996,piskunovSME2017}. SME uses a grid of MARCS models\footnote{Available at \url{(marcs.astro.uu.se)}.} \citep{gustafsson2008} with spherical symmetry for \logg $< 3.5$, which is applicable to the \fstar\ and the bulge stars analysed in the Jönsson-series. The atomic data comes from the Gaia-ESO line list version 6 \citep{heiter2021A&A...645A.106H}.

In order for SME to produce a synthetic spectrum, it requires a line list, model atmospheres, defined spectral segments, as well as spectral line- and continuum masks. The masks are defined manually around the spectral line of interest, which in \cite{jonssona2017A&A...598A.100J,jonsson2017b,lomaeva2019,Forsberg2019,forsbergsubmitted} has been shown to be crucial in order to get the high-precision abundances they report. 

\subsection{Europium}
\label{sec: abundance determination}
For europium (Eu, Z\,=\,63), we use the Eu II 6645 Å spectral line for the abundance determination, with the atomic data from the Gaia-ESO (GES) line list version 6 \citep{heiter2021A&A...645A.106H}. They classify the 6645 Å line as a \textit{Yes}/\textit{Uncertain}, meaning that the $\log(gf)$-value is of high quality, however that there could be some uncertain blends in the line. They recommend the line to be used only for giant stars with high S/N spectra, which is the case for the \citeauthor{zoccali2006}-spectrum of the \fstar  used here. 

Eu has both hyperfine structure and two stable isotopes in the Sun. The hyperfine splitting is included in the GES line list, whereas the isotopic shift can not be resolved and is not included.

This line has been used with the same analysis as described here to determine the Eu abundance for close to 300 disc K-giants in \citet{Forsberg2019}. Their abundance trend is tightly correlated with [Fe/H], giving confidence in the quality of our analysis presented here. However, we do note that the [Eu/Fe] abundances seems to be systematically too high in the disc stars in \citet{Forsberg2019}, which might be explained by the possible systematic uncertainty in their \logg as reported in \citet{jonssona2017A&A...598A.100J,jonsson2017b} \citep[when compared to Gaia Benchmark stars in][]{jofre2014A&A...564A.133J,jofre:2015,heiter2015A&A...582A..49H}. Therefore we systematically lower the overall disc abundances with 0.10\,dex such that the thin disc trend passes through the Solar value, as can be seen in Fig.~\ref{figure: [Eu/Fe] bulge, disk and halo}. For the same reasons, the [La,Ce/Fe] values are also systematically lowered, which can be seen in Table~\ref{table:norm abundances}.

Nonetheless, the [Eu/Fe] value for the \fstar  from their study remains high, which is confirmed by determining the abundance using the \citet{lecureur2007,zoccali2008,ryde2010,johnson2014AJ....148...67J} stellar parameters as well, that do not report any systematics in their parameters, indicating the Eu-content to indeed be high in the \fstar.

\subsection{Molybdenum}
For molybdenum (Mo, Z\,=\,42), we use the Mo I 6030 Å spectral line for abundance determination, which is very weak in dwarf stars but usable in giant stars \citep{forsbergsubmitted}. In \citet{heiter2021A&A...645A.106H} the line is reported to be \textit{Yes}/\textit{Yes}, having well-known atomic data and being unblended. Mo does not have hyperfine structure, but has seven stable isotopes in the Sun. Similarly to Eu, this isotopic splitting is not included in the linelist, and we determine the abundance using the five set of stellar parameters in Table \ref{table: stellar parameters}.

The manually defined line- and continuum masks can be seen in Fig.~\ref{fig: continuum and line segments of Mo, Eu}. Both the spectrum of the \fstar\ and the synthetic spectrum can be seen in Fig.~\ref{fig: spectrum of Mo, Eu}, with a $\pm 0.1$\,dex uncertainty.

\begin{figure*}[ht!]
   \centering
   \includegraphics[width=\hsize]{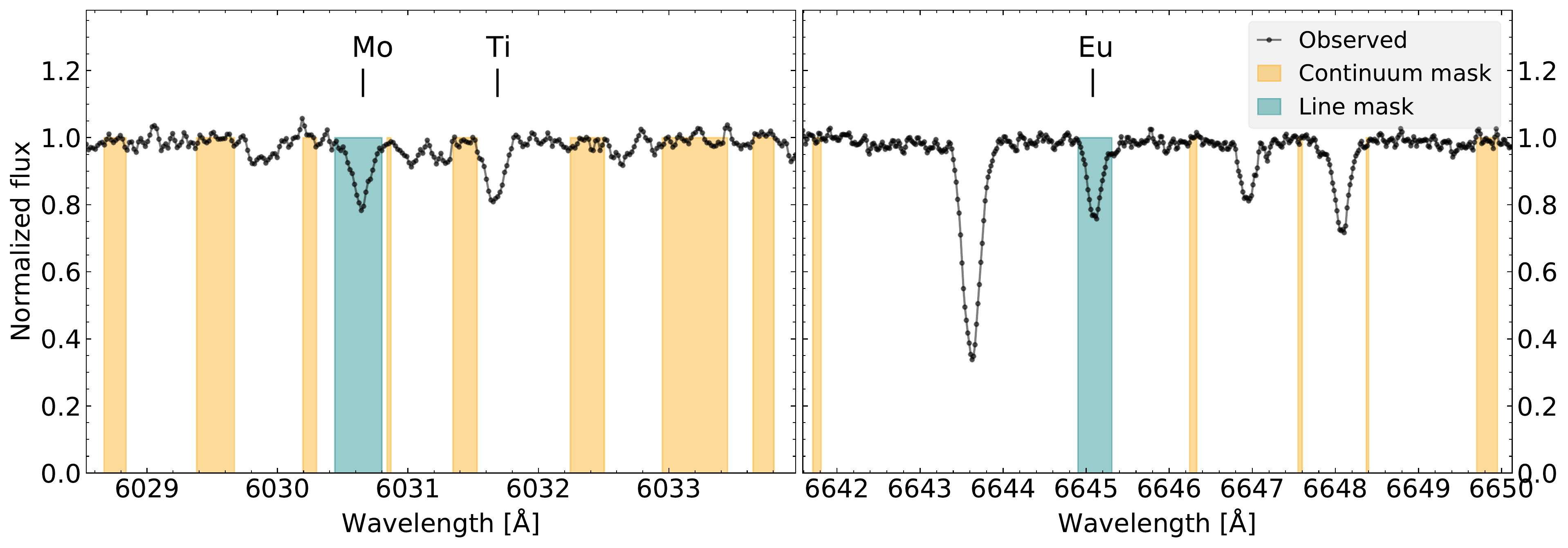}
      \caption{The observed spectrum of the \fstar\ in black with a SNR of 65. The line mask placements for the Mo (left) and Eu (right) line can be seen in turquoise and the continuum placements in yellow.}
      \label{fig: continuum and line segments of Mo, Eu}
\end{figure*}

\begin{figure*}[ht!]
   \centering
   \includegraphics[width=\hsize]{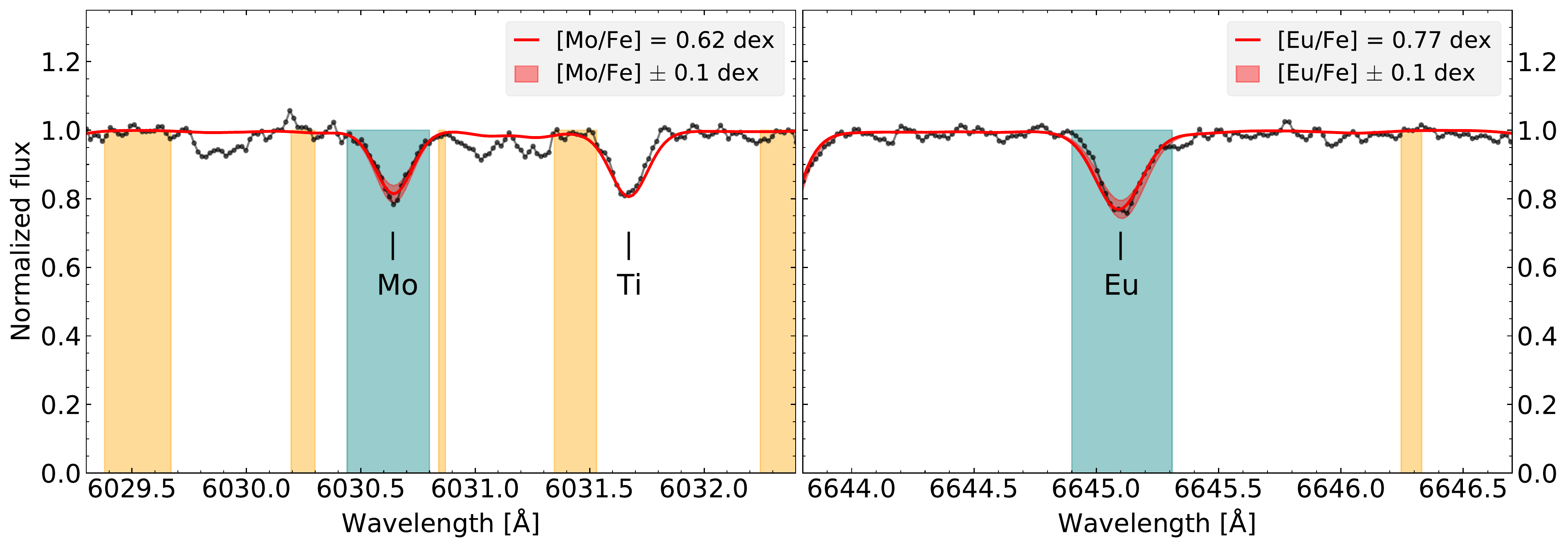}
      \caption{Same as Fig.~\ref{fig: continuum and line segments of Mo, Eu}, with the synthetic spectra seen in red. The best fit is shown with a solid red line and the $\pm$ 0.1\,dex with a shaded red. Note that the wavelength region is zoomed in compared to Fig.~\ref{fig: continuum and line segments of Mo, Eu} to better show the fit of the synthetic spectrum.}
      \label{fig: spectrum of Mo, Eu}
\end{figure*}

\begin{table*}
    \caption{Comparison of the stellar parameters (Col.~2-5) for the \fstar\ from five different spectroscopic studies (Col.~1). Note that the \citet{ryde2010} stellar parameters are determined from an infrared, H-band spectrum of the \fstar. We calculate the [Mo/Fe] and [Eu/Fe] (column 6-7) using the methodology as described in Sect.~\ref{sec: analysis if spectrum}, and use A(Mo)$_\odot = 1.88$ and A(Eu)$_\odot = 0.52$ from \citet{grevesse2015A&A...573A..27G}.}         
    \label{table: stellar parameters}  
    \centering     
    \begin{tabular}{c | c | c | c | c || c | c}  
    \hline\hline     
    Reference&\teff&\logg &\feh&\vmic &[Mo/Fe]&[Eu/Fe] \\
    & [K]& [dex]&[dex]& [km s$^{-1}$]&[dex]&[dex] \\
    \hline \hline 
    \citet{lecureur2007} & 4400 & 1.8 & -0.62 & 1.4 & 0.73 & 0.85 \\
    \citet{zoccali2008} & 4350 & 1.7 & -0.65 & 1.4 & 0.68 & 0.83 \\
    \citet{ryde2010} & 4250 & 1.5 & -0.69 & 1.4 & 0.58 & 0.81 \\
    \citet{johnson2014AJ....148...67J} &  4425 & 1.65 & -0.58 & 1.70 & 0.68 & 0.74 \\
    \citet{jonsson2017b} &  4287  & 1.79  & -0.67 & 1.46 & 0.62 & 0.78 \\
    \hline \hline   
    \end{tabular}
    \tablefoot{Note that Eu has been decreased by 0.10 dex, since \citet{Forsberg2019} report possible [Eu/Fe] to possibly be systematically too high due to systematics in the stellar parameters. We lower it such that their overall thin disc, as seen in Fig.~\ref{figure: [Eu/Fe] bulge, disk and halo}, goes through the solar value, also as described in Sect.~\ref{sec: abundance determination}}
\end{table*}

\begin{table}
\caption{Determined absolute abundances A(X) and solar- and metallicity normalised abundances [X/Fe] of the \fstar. The Jönsson series, \citet{jonsson2017b,lomaeva2019,Forsberg2019} and \citet{forsbergsubmitted} uses the solar values in the series of \citet{scott..a..2015A&A...573A..25S,scott..b..2015A&A...573A..26S,grevesse2015A&A...573A..27G}. For elements not determined in that series, we complement with values from \citet{ryde2010,johnson2014AJ....148...67J,vanderswaelmen}. The C and N abundances are from molecular CO and CN \citep{ryde2010}.} 
\label{table:norm abundances}      
\centering          
\begin{tabular}{l | c | c | l }    
\hline\hline   
Species & A(X) & [X/Fe] & Reference \\ \hline \hline
C     &  7.63 & -0.10  & \citet{ryde2010} \\
N     &  7.34 &  0.21  & \citet{ryde2010} \\
O I   &  8.47 &  0.48  & \citet{jonsson2017b} \\
Na I  &  5.67 & -0.08  & \citet{johnson2014AJ....148...67J}  \\
Mg I  &  7.27 &  0.47  & \citet{jonsson2017b} \\
Al I  &  6.16 &  0.27  & \citet{johnson2014AJ....148...67J} \\
Si I  &  7.21 &  0.24  & \citet{johnson2014AJ....148...67J} \\
Ca I  &  5.88 &  0.25  & \citet{jonsson2017b} \\
Sc II &  2.57 &  0.08  & \citet{lomaeva2019} \\
Ti I  &  4.43 &  0.19  & \citet{jonsson2017b} \\
V I   &  3.34 &  0.12  & \citet{lomaeva2019} \\
Cr I  &  4.48 & -0.11  & \citet{lomaeva2019} \\
Co I  &  4.40 &  0.14  & \citet{lomaeva2019} \\
Ni I  &  5.57 &  0.04  & \citet{lomaeva2019} \\
Zr I  &  2.35 &  0.43  & \citet{Forsberg2019} \\
Mo I  &  1.87 &  0.62  & \citet{forsbergsubmitted}  \\
Ba II &  2.33 &  0.78  & \citet{vanderswaelmen}  \\
La II &  0.93 &  0.42* & \citet{Forsberg2019} \\
Ce II &  1.11 &  0.22*  & \citet{Forsberg2019} \\
Nd II &  1.15 &  0.32  & \citet{vanderswaelmen}  \\
Eu II &  0.73 &  0.78* & \citet{Forsberg2019} \\
\hline \hline 
\end{tabular}
\tablefoot{Note that La, Ce and Eu has been decreased by 0.05, 0.01, and 0.10 dex, as described in Sect.~\ref{sec: abundance determination}.}
\end{table}

\section{Bulge membership}
\label{sec: bulge membership investigation}
A majority of the r-process enriched stars are metal-poor, [Fe/H]\,$\leq -1.5$\,dex, and confined to the Galactic halo \citep{holmbeck2020ApJS..249...30H}. In order to investigate whether the \fstar\ and the \jstar\ are likely to be confined to the Galactic bulge or visiting halo-stars, we estimate their line-of-sight distances and, using, their proper motions \citep[from Gaia EDR3,][]{gaiacollaboration2016A&A...595A...1G,gaiacollaboration2021A&A...649A...1G} we estimate their orbits using \texttt{galpy} \citep{galpybovy2015ApJS..216...29B}. The line of sight position of both stars can be seen in Fig.~\ref{fig:bulgefields}, where the \fstar\ is closer to the Galactic plane, compared to the \jstar.


\subsection{Distance estimates}
\label{sec: distance estimation}
We calculate spectrophotometric distances for the two stars where we incorporate the stellar properties $\rm T_{eff}$, $\log(g)$, [Fe/H] and the 2MASS JHKs photometry into the spectrophotometric method described in \citet{Rojas-Arriagada2017A&A...601A.140R,rojasarriagada2017A&A...601A.140R}. For the stellar parameters, we use the \citet{jonsson2017b} parameters for the \fstar, and \citet{johnson2013ApJ...775L..27J} parameters for the \jstar. Each parameter point in the parameter space $\rm T_{eff}$, $\log(g)$, [Fe/H] is compared to a set of theoretical isochrones for each star. We use the PARSEC isochrones \citep{bressan2012MNRAS.427..127B,marigo2017ApJ...835...77M} spanning ages from 1 to 13\,Gyr in steps of 1\,Gyr and metallicities from -2.2 to +0.5 in steps of 0.1\,dex. Note that the PARSEC isochrones do not take $\alpha$-elements into account.

A number of extra multiplicative weights are defined to account for the evolutionary speed of the points along the isochrones and for the initial mass function. Using these weights, the most likely absolute magnitudes (MJ, MH, MKs) of the observed stars can be computed as the weighted mean or median of the theoretical values of the whole set of isochrone points. The computed absolute magnitudes are then compared to the observed photometry, allowing us to estimate the line-of-sight reddening and distance modulus. These distances were extensively tested and used within the APOGEE survey \citep{apogee2017} \citep[see e.g.][]{Rojas-Arriagada2020,Zasowski2019}. 

For the \fstar\ we derive a heliocentric distance of $\rm 7.2 \pm 0.3\,kpc$ and for the \jstar\ $\rm 11.8 \pm 1.0\,kpc$. The uncertainties for the distances are estimated using the reported uncertainties of the photometry and stellar parameters. These distances indicate that the \fstar\ belongs to the Galactic bulge, while the large distance of the \jstar\ suggests, together with its low metallicity of $\rm[Fe/H]= -1.69$\,dex that it originates from either the Galactic halo or thick disc. This hypothesis was indeed already suggested by \cite{johnson2013ApJ...775L..27J}. 

In addition, we have calculated the distances using the Gaia DR3 photometry in the G-band \citep{Kordopatis2022} for these two stars in order to check the consistency. A distance of $\rm 7.4 \pm 1.0\,kpc$ has been obtained for the \fstar\ and of $\rm 12.2 \pm 1.5\,kpc$ for the \jstar, respectively. This strengthen our argument that indeed the \jstar\ is a visiting halo/inner halo or thick disc star, while the \fstar\ belongs to the Galactic bulge.

\subsection{Modelling of orbits}
\label{sec: orbital method}
To further investigate the bulge membership, we explore possible orbits of the stars using \texttt{GalPy} \citep{galpybovy2015ApJS..216...29B}. We adopt the Milky Way like \texttt{MWPotential2014}-potential \citep{galpybovy2015ApJS..216...29B} which includes an axisymmetric disc, a halo and a spherical bulge. To that we add the \citet{dehnen2000AJ....119..800D}-bar potential \citep[which in \texttt{GalPy} has been generalised to 3D following][]{monari2016MNRAS.461.3835M}.

We integrate the orbits up to 10 Gyr for both stars, using positions in RA+Dec; proper motions \citep{gaiacollaboration2021A&A...649A...1G}; distances as derived above; and radial velocities determined in \citet{johnson2013ApJ...775L..27J, jonsson2017b}. We also calculate the orbits using the uncertainties for distances and proper motions to check whether the orbits remain similar. The data used can be seen in Table \ref{table: positonal and kinematic data}. As a sanity check, we also calculate the orbits using the code \texttt{AGAMA} \citep{agama2019} where we implement a combined \texttt{MWPotential2014} and \citet{Launhardt2002}-inner bulge potential. The code gives similar results as the ones presented for \texttt{GalPy}.




\begin{table}[t]
\caption{RA+Dec and proper motions from Gaia EDR3 \citep{gaiacollaboration2016A&A...595A...1G,gaiacollaboration2021A&A...649A...1G}. Distance estimates, as described in Sect.~\ref{sec: distance estimation}. The JHK photometry comes from 2MASS \citep{cutri2003yCat.2246....0C}. The radial velocities comes from \citet{johnson2011ApJ...732..108J} and \citet{jonsson2017b}, respectively for the \jstar\ and \fstar.} 
\label{table: positonal and kinematic data}      
\centering                          
\begin{tabular}{c | c | c }        
\hline\hline                 
Properties & \jstar & \fstar \\    
\hline \hline 
(l,b) [deg] & (-1.0, -8.4) & (5.2, -2.8) \\
RA [deg] & 274.438 & 272.10 \\
Dec [deg] & -33.890 & -25.812 \\
J mag & 11.686 & 12.255 \\
H mag & 11.034 & 11.395      \\       
K mag & 10.955 & 11.130        \\
Distance [kpc] & $\rm 11.8 \pm 1.0$ & $\rm 7.2 \pm 0.3$ \\
pm(ra) [mas/yr] & -3.734$\pm 0.023$ & -0.304 $\pm 0.052$ \\
pm(dec) [mas/yr] & -9.512 $\pm 0.018$ & -2.601 $\pm 0.036$\\
RV [km/s] & -16 & -36 \\ 
\hline                                   
\end{tabular}
\end{table}

\vspace{-5pt}
\section{Results and discussion}
\label{sec: results and discussion}
We present the abundances of the \fstar\ and compare with both the \jstar\ and the overall bulge sample as determined in our series of papers. We will show the orbits, discuss these and the distances, and consider implications for the formation history of the bulge.

\vspace{-5pt}
\subsection{Abundances}
We re-analyse the spectrum of the \fstar\ and determine both the [Eu/Fe] and [Mo/Fe] abundance using the stellar parameters reported in \citet{lecureur2007,zoccali2008, ryde2010,johnson2014AJ....148...67J,jonsson2017b}, which can be seen in Table \ref{table: stellar parameters}. 

From these, the \fstar\ has a mean [Eu/Fe] = 0.80\,dex, based on the five different stellar parameter determinations, pointing at the Eu-content being high in the \fstar, independently on the stellar parameters. All of the values are [Eu/Fe]\,$\geq +0.74$\,dex, which classifies it as a, relatively metal-rich, r-II star \citep[following the classification in the R-process Alliance,][]{holmbeck2020ApJS..249...30H}. In Fig.~\ref{figure: [Eu/Fe] bulge, disk and halo} we show [Eu/Fe] vs [Fe/H] for the bulge stars in \citet{Forsberg2019} and \citet{johnson2012ApJ...749..175J}, which both include the \fstar\ and the \jstar, respectively. We also plot the typical local disc abundances \citep{Forsberg2019} and stars from the R-process Alliance \citep{holmbeck2020ApJS..249...30H}. Note that even though the [Eu/Fe]-abundances from \cite{Forsberg2019} has been systematically lowered with 0.10\,dex, as described in Sect. \ref{sec: abundance determination}, both the \fstar\ and the \jstar\ lie above the \citet{holmbeck2020ApJS..249...30H}-classification boundary for an r-II star, and are high in [Eu/Fe] as compared to the bulge samples.

   \begin{figure}[ht!]
   \centering
   \includegraphics[width=\hsize]{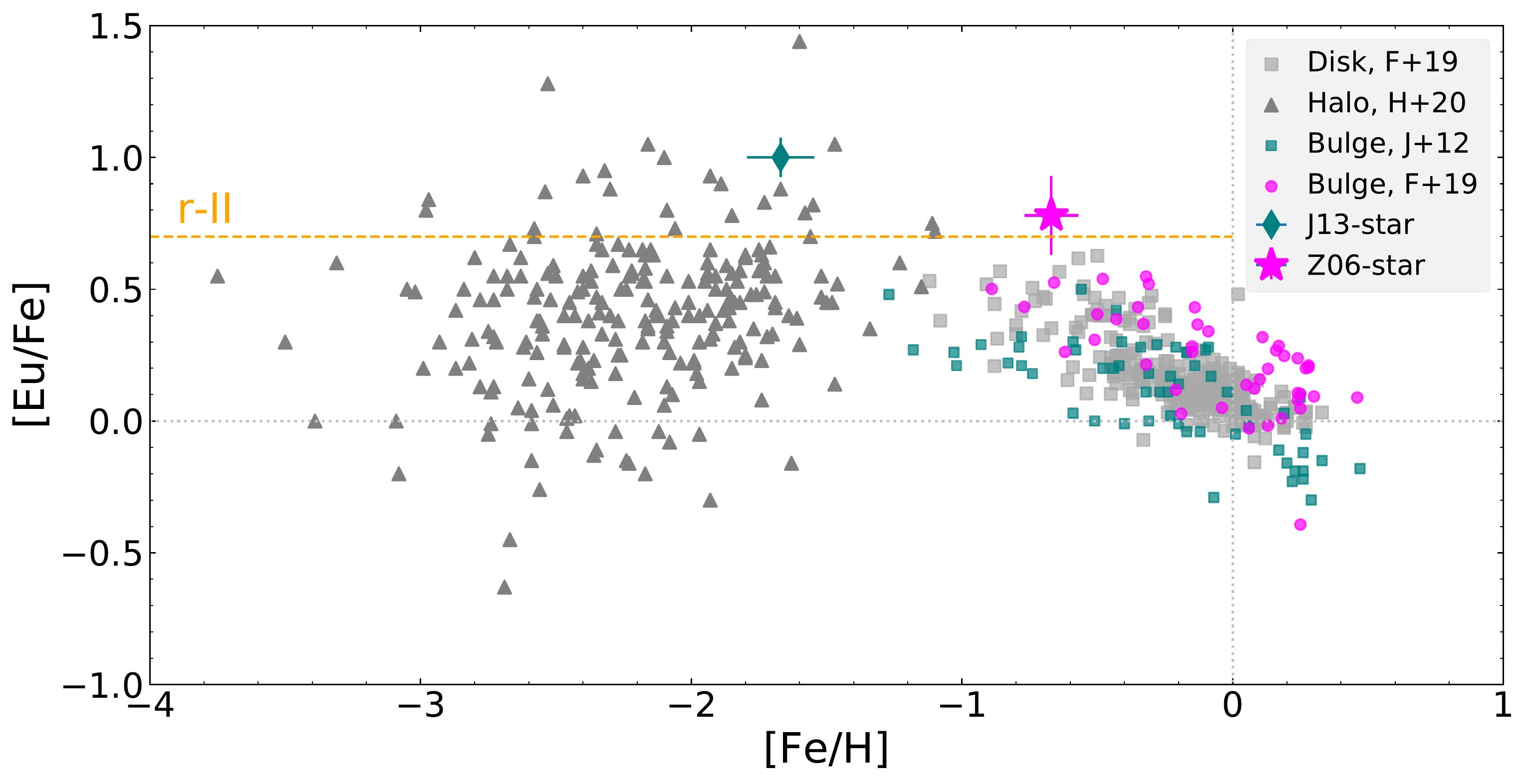}
      \caption{[Eu/Fe] over [Fe/H] for the \fstar\ (magenta star) and the \jstar\ (teal diamond). For reference, we plot the bulge (magenta circles) and disc stars (grey squares) from \citet{Forsberg2019} as well as the bulge stars from \citet{johnson2012ApJ...749..175J}. The abundances of the halo stars come from the r-process alliance sample \citep[dark grey triangles][]{holmbeck2020ApJS..249...30H}. The orange line indicates the 'r-II' limit as defined in the r-process alliance \citep{holmbeck2020ApJS..249...30H}. Note that the abundances from \citet{Forsberg2019} has been systematically lowered with 0.10\,dex, as described in Sect. \ref{sec: abundance determination}.}
         \label{figure: [Eu/Fe] bulge, disk and halo}
   \end{figure}
  
The [Mo/Fe] abundance for the \fstar\ is high with $0.62$\,dex from the \citet{forsbergsubmitted} study, and an average of $0.66$\,dex when determined using the five different stellar parameter studies (Table \ref{table: stellar parameters}). In Fig. \ref{fig: [Mo/Fe] bulge, disk and halo} we show the [Mo/Fe] abundances for the local disc and the bulge \citep{forsbergsubmitted}, including the \fstar. We have also added the \jstar\ and molybdenum abundances from disc and halo studies \citep{peterson2013ApJ...768L..13P,hansen2014A&A...568A..47H,roederer2014...313...AJ....147..136R,mishenina2013A&A...552A.128M}.

\begin{figure}[ht]
   \centering
   \includegraphics[width=\hsize]{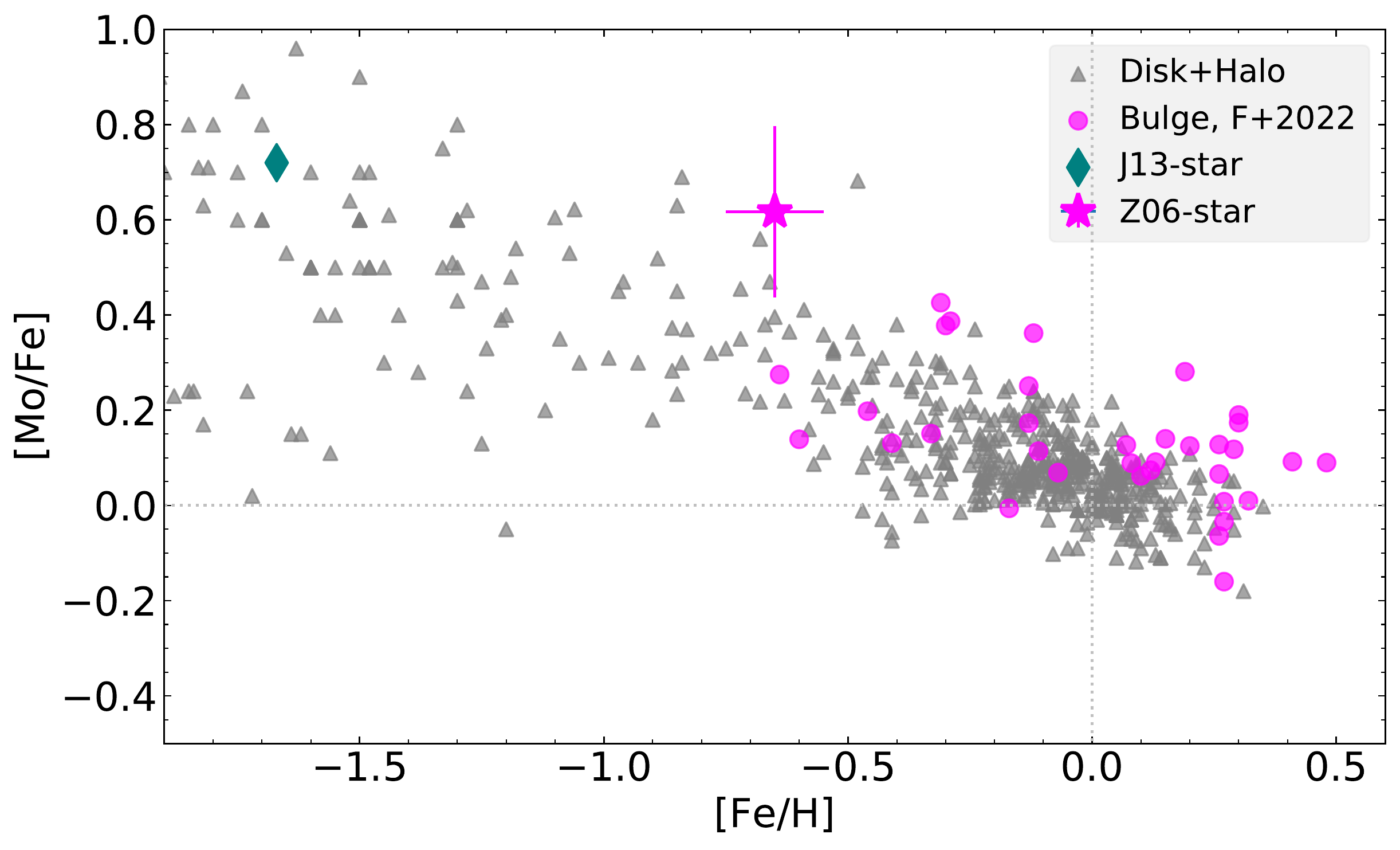}
      \caption{[Mo/Fe] over [Fe/H] for the \fstar\ (magenta star) and \jstar\ (teal diamond). We also show the bulge in [Mo/Fe] from \citet[][magenta circles]{forsbergsubmitted}. The disc and halo with grey triangles are from \citep{peterson2013ApJ...768L..13P,hansen2014A&A...568A..47H,roederer2014...313...AJ....147..136R,mishenina2019,forsbergsubmitted}.}
    \label{fig: [Mo/Fe] bulge, disk and halo}
\end{figure}

\begin{figure}[ht!]
   \centering
   \includegraphics[width=\hsize]{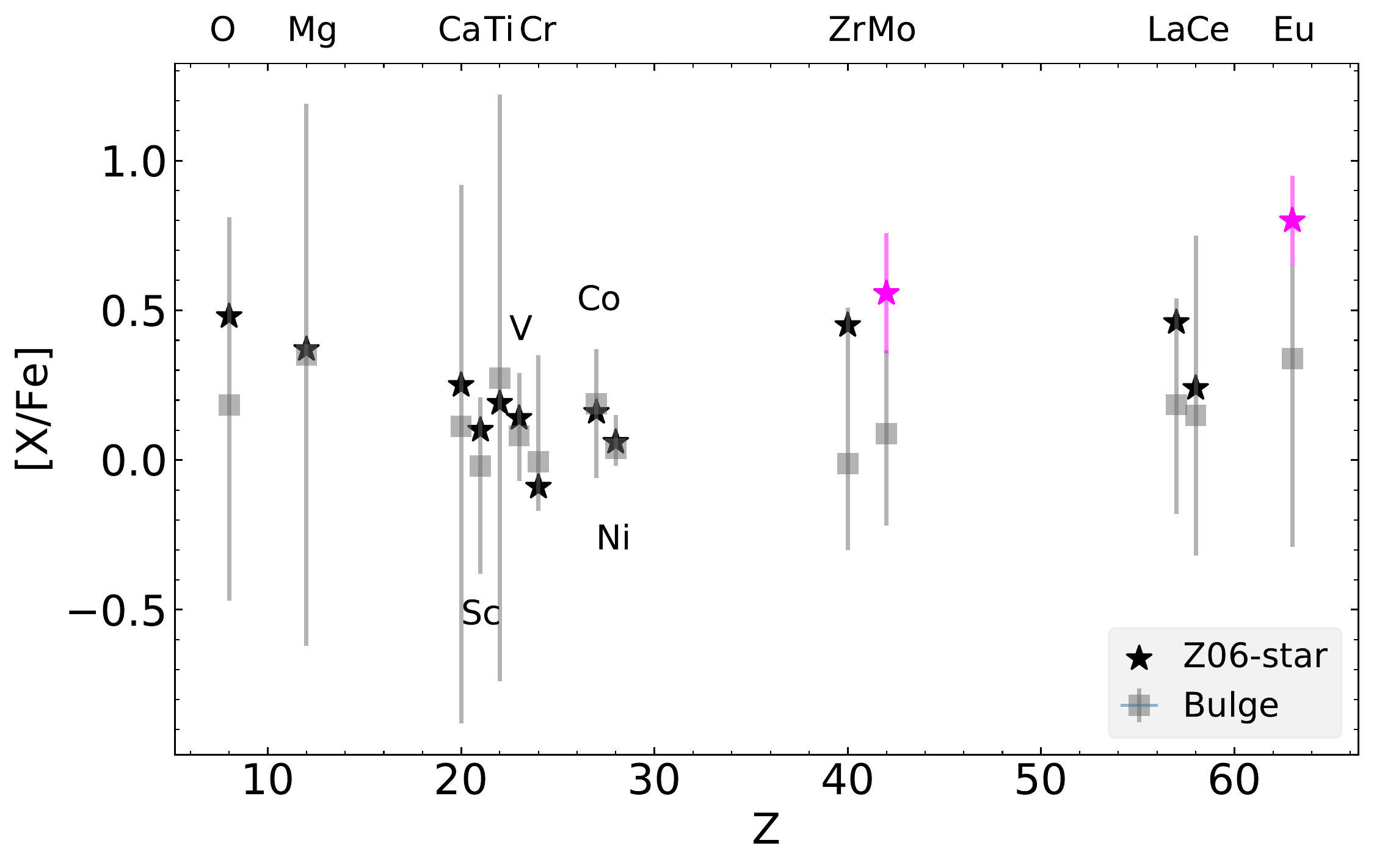}
   \includegraphics[width=\hsize]{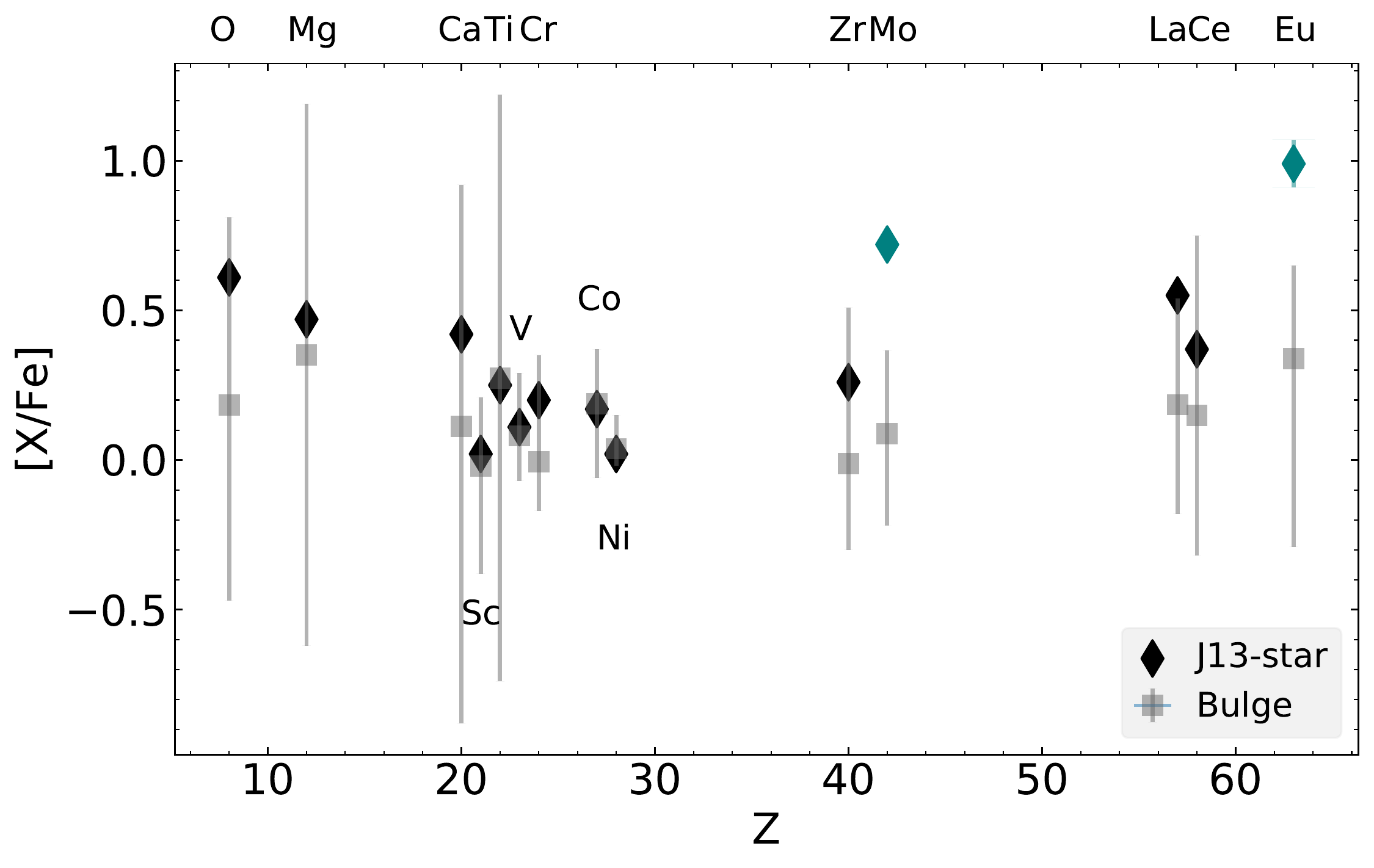}
      \caption{[$X$/Fe] over Z for the bulge abundances (grey) presented in the Jönsson-series \citep{jonsson2017b,lomaeva2019,Forsberg2019,forsbergsubmitted} for the \fstar\ (upper plot, indicated with a star-symbol and the estimated uncertainties) and \citet{johnson2013ApJ...775L..27J}-star (lower plot, indicated with a diamond). The bulge abundances are indicated by the mean (grey box) with the highest and lowest value (the \fstar\ excluded) indicated by the bar, showing the spread in abundance in the bulge sample. To facilitate the reading of the plot, the element corresponding to a certain Z has been indicated. The r-process elements Mo and Eu has been indicated using magenta (\fstar) and teal (\jstar).}
         \label{fig: abundances with Z}
\end{figure}

In order to compare the stars with each-other, in Fig.~\ref{fig: abundances with Z} we compare their [$X$/Fe] in with the overall bulge abundances as measured from the Jönsson-series \citep{jonsson2017b,lomaeva2019,Forsberg2019,forsbergsubmitted}. We see that both the \fstar\ and the \jstar\ are enhanced in [Mo/Fe] and [Eu/Fe]. Additionally, both stars are on the higher side in the s-process elements [Zr,La/Fe]. This pattern of high abundance in heavier elements is similar to that of the prototypical r-process enhanced metal-poor halo stars BD +17$^{\circ}$3248 \citep{cowan2002ApJ...572..861C}, CS 22892-052 \citep{sneden2003ApJ...591..936S} and HD 222925 \citep{roederer2022}.

Europium has a close to pure r-process origin with 96\,\% r-process contribution at solar metallicites. Molybdenum has a more complex cosmic origin, as is composed of seven stable isotopes, with both s-, r- or/and p-process origin. The r-process contribution varies a bit in the literature, from 27\,\% to 36\,\% at solar metallicities \citep[][respectively]{prantzos020MNRAS.491.1832P,bisterzo2014ApJ...787...10B}. Mo has a roughly 25\,\% origin from the p-process, which have several suggested cosmic sites and mechanisms behind it. It is likely to take place in proton-dense and/or explosive environments, which points toward fast timescales, such as core-collapse SNe, events linked to SNe type Ia, or mass-accreting neutron stars \citep[see e.g.][]{rauscher2013,forsbergsubmitted}. At the low metallicites of $-0.65$ and $-1.67$\,dex for the stars, the s-process, which takes place in AGB-stars, is less likely to  contribute to the production the Mo \citep{karakas2014,cescutti2015A&A...577A.139C,cescutti2022Univ....8..173C}. As such, at low metallicities Mo is effectively an r-process and p-process element. The same applies to Zr and La, which at lower metallicities have a dominating production from the r-process.

To decipher if the high neutron-capture abundances has an r- or s-process origin, we consider the [s/r]-ratio. Both the \fstar\ and the \jstar\ have rather low [La/Eu]-ratio of $-0.36$ and $-0.44$\,dex, respectively. Additionally, the [Ce/Eu]-ratio\footnote{La and Ce has a roughly 75\,\% and 85\,\% origin from the s-process at solar metallicities, respectively \citep{bisterzo2014ApJ...787...10B,prantzos020MNRAS.491.1832P}.} is –0.56 and $-0.62$\,dex respectively for the \fstar\ and \jstar. This supports a significant contribution from the r-process in the enrichment of these two stars, and less significant contribution from the s-process \citep[which indeed is the case for many bulge stars, see e.g.][]{mcwilliam2016PASA...33...40M}.

Although many stars with high r-process enhancements are known \citep{cowan2002ApJ...572..861C,sneden2003ApJ...591..936S,sneden2008ARA&A..46..241S,Christlieb2004A&A...428.1027C,  holmbeck2020ApJS..249...30H}, it is rare to find one at the comparatively high metallicities of the \jstar\ and the \fstar. It should be noted that the metallicity distribution in the bulge varies with latitude, where the metallicity is higher towards the plane \citep[see e.g.][]{johnson2020}. As such, the \fstar\ is metal-poor relative to its neighbouring stars. The same applies for the \jstar, where at the latitudes of $b = -8^{\circ}$, lower metallicity stars are more common. Nonetheless, the relatively high metallicities of the stars (for being r-enriched stars) may likely be explained by a combination of iron-producing SNe driving the metallicity up, with multiple r-process events \citep[NSM, MRSNe,][]{matteucci2014MNRAS.438.2177M,kobayashi2020ApJ...900..179K} and Galactic mixing that inhibits otherwise high [$X$/Fe]-abundances. A more robust interpretation awaits surveys where we might answer whether the \fstar\ is indeed unusual or whether more r-enhanced bulge giants are found near [Fe/H]$\sim -0.5$\,dex, perhaps relics of an early formation epoch in the bar. 

\subsection{The orbits}
\begin{figure}[ht]
   \centering
     \includegraphics[width=\hsize]{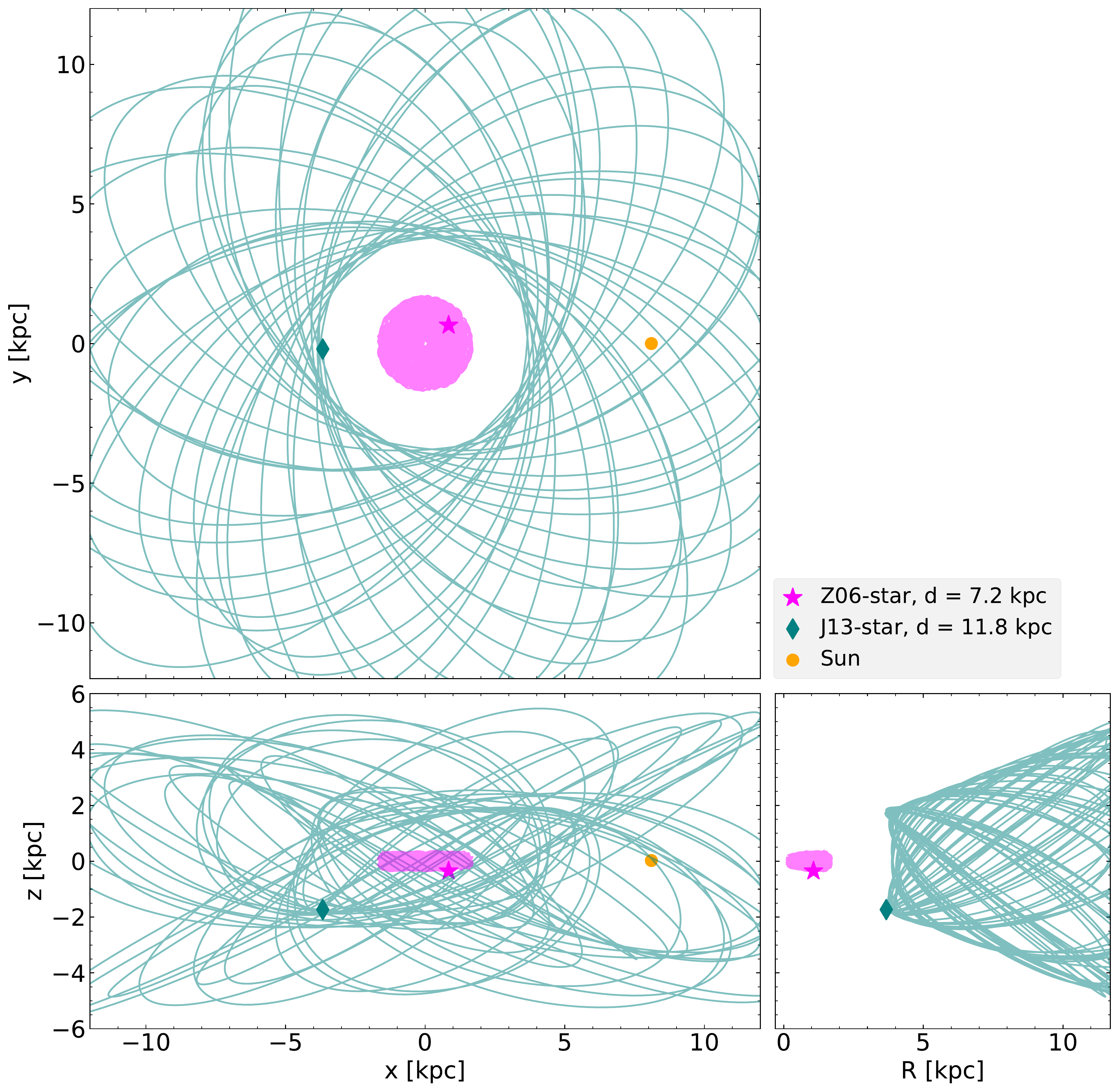}
        \caption{Orbit for the \jstar\ (teal) and the \fstar\ (magenta) plotted in x-y (top), x-z (bottom left) and R-z (bottom right) plane. The orbits are integrated over 10 Gyr and the position of the stars in their orbits are indicated, as well as the position of the Sun (yellow dot). This is the nominal case for the distances ("d"), which are indicated in the legend, and seen in Table \ref{table: positonal and kinematic data}.}
         \label{figure: Orbitals R-z} 
   \end{figure}

\begin{figure}[ht]
   \centering
     \includegraphics[width=\hsize]{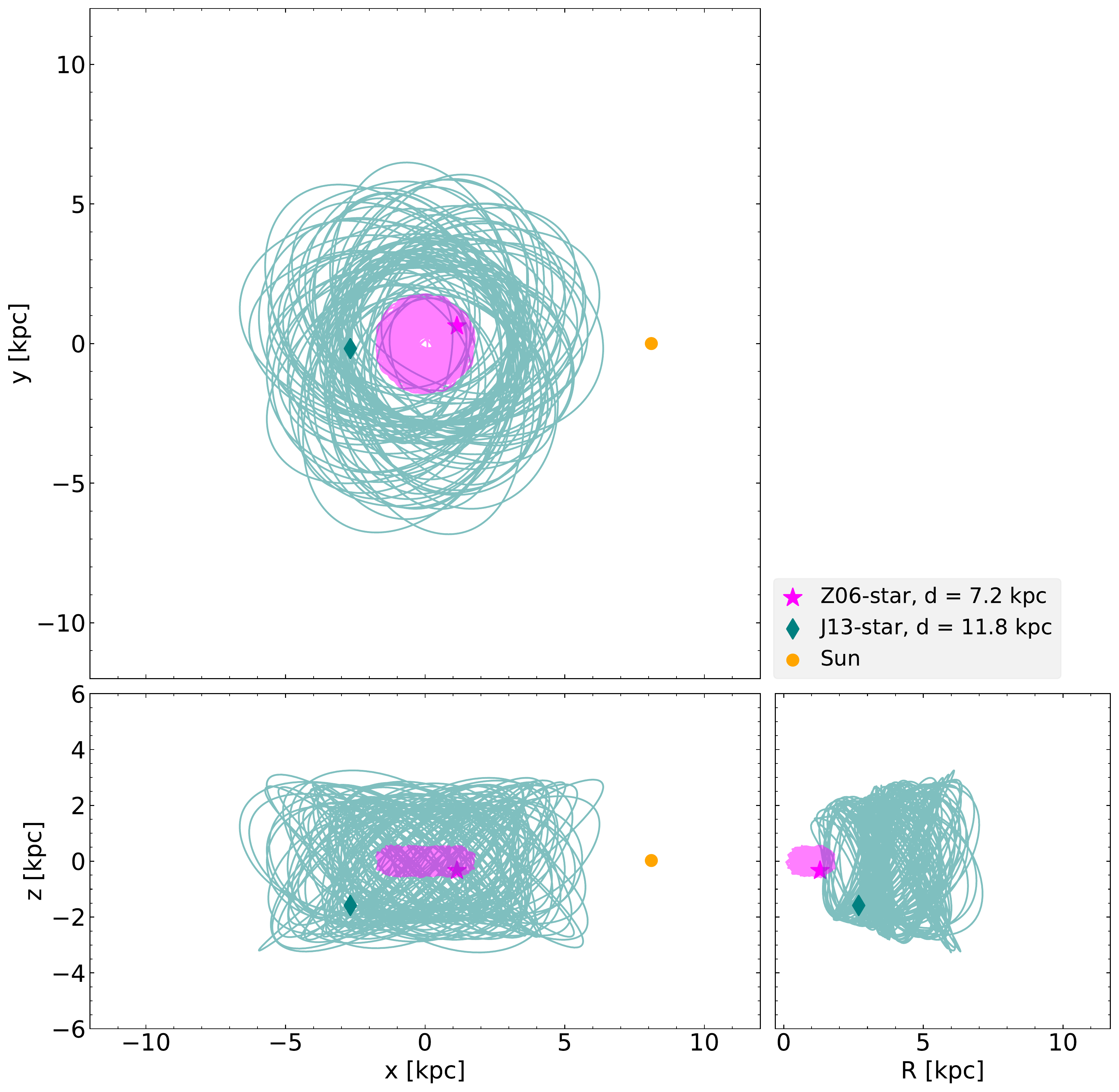}
     \includegraphics[width=\hsize]{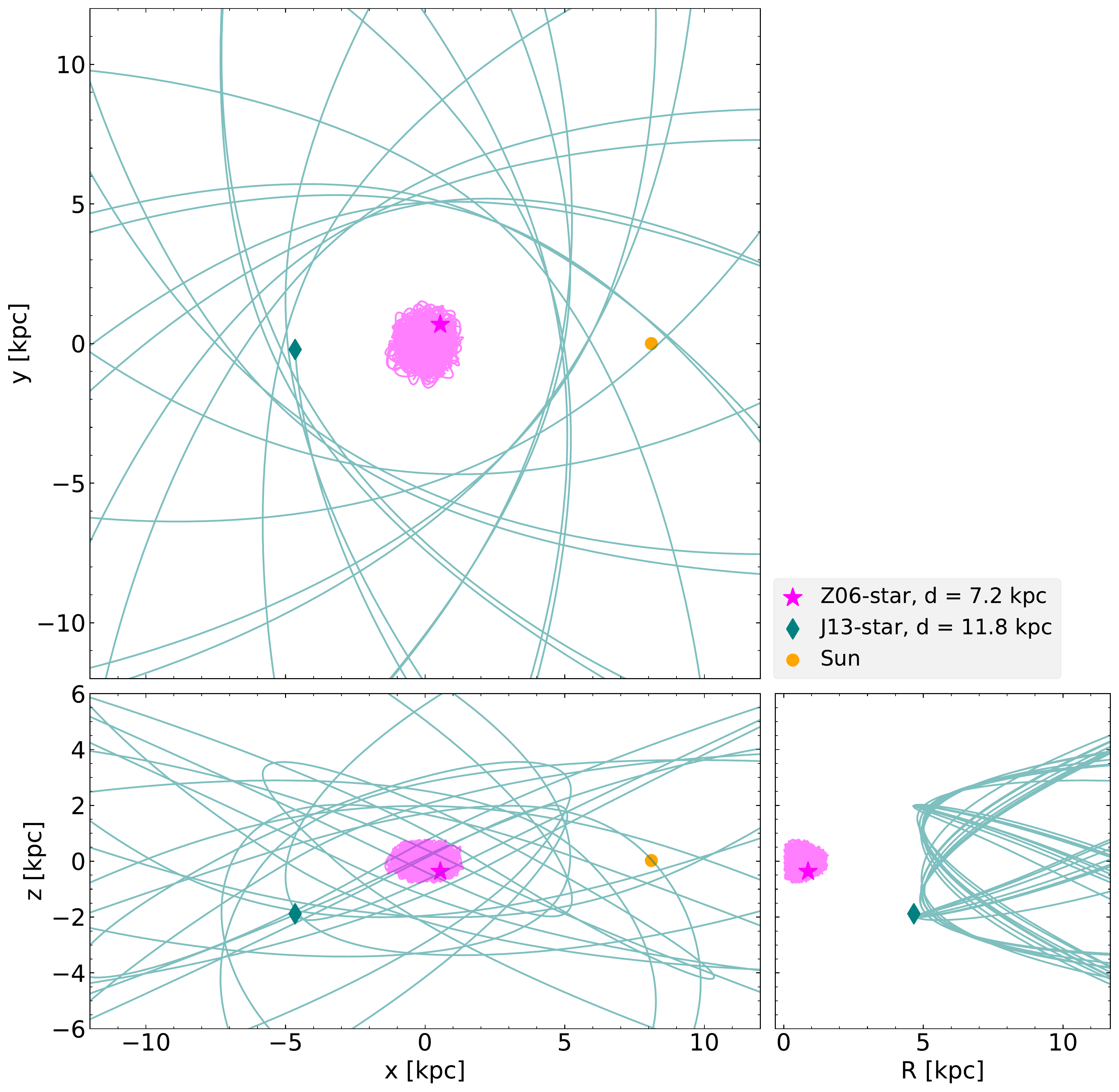}
        \caption{Same as Fig.~\ref{figure: Orbitals R-z} but with the low- and high distance ("d") uncertainties (top and bottom, respectively), as indicated in the legend. Similarly as the nominal case, we integrate the orbits for 10 Gyr.}
         \label{figure: Orbitals R-z errors} 
   \end{figure}

In Fig.~\ref{figure: Orbitals R-z} we show the orbits for both stars. The \jstar\ makes clear deviations to both high $|\mathrm{z}| > 5$\,kpc and even passes outside of the Sun's orbit in x-y plane. We also estimate the orbital parameters taking into account the derived uncertainties in distance and proper motion, and find that the distance is the parameter that makes the largest difference, and the proper motion is less important when the uncertainty in the distance is large. In Fig.~\ref{figure: Orbitals R-z errors} we show the orbits with the low and high distance estimates (as taken from Table \ref{table: positonal and kinematic data}) using the nominal proper motions. The \fstar\ always remains in the bulge, given its smaller uncertainty in distance, whereas the \jstar\ is heavily affected by its distance. For the lower distance of the \jstar, it extends to $|\mathrm{z}| \sim 3$\,kpc, and seems to trace the bar potential adopted in the \texttt{GalPy}-setup (see Sect.~\ref{sec: orbital method})

As for the distances of the stars, we do note that both stars have distance estimates in \citet{bailerejones2021AJ....161..147B} which uses a Bayesian approach and a prior model of the Galaxy. The distances are estimated to $7409^{+1369}_{-972}$\,pc and $6839^{+2375}_{-1252}$\,pc for the \jstar\ and the \fstar, respectively\footnote{Note that this is the geometric distance as measured by \citet{bailerejones2021AJ....161..147B}, the photogeometric, which combines the geometric with G-magnitude and BP-RP colours, gives $6316^{+1257}_{-747}$ pc and $6559^{+1781}_{-1372}$ pc, respectively for \jstar\ and \fstar. In general, the photogeometric distance extend to larger distances and tend to have smaller uncertainties.}. The reader are referred to \citet{bailerejones2021AJ....161..147B} for more details on the distance measurements. 

Similarly, the StarHorse code \citep[SH,][]{Queiroz2018MNRAS.476.2556Q}, which also uses a Galactic prior, has distance estimates and stellar parameters to the stars using Gaia EDR3 and photometric catalogues in \citet{anders2022A&A...658A..91A}. They give the distances to 6.1\,kpc and 7.4\,kpc, respectively for the \jstar\ and the \fstar. 

These distance measurements puts the \jstar\ within the bulge, however, we also do note that the metallicity output from SH is very high, $-0.38$\,dex, compared to $-1.67$\,dex from the spectroscopic measurements \citep{johnson2013ApJ...775L..27J}. Similarly, they estimate a 0.18\,dex metallicity for the \fstar, indicating large uncertainties when determining metallicities for very distant stars, which propagate to the estimated distances. Additionally, the parallax error in Gaia EDR3 for this star is as high as $1/4^{\mathrm{th}}$ of the parallax (\texttt{parallax\textunderscore over\textunderscore error} of 3.9), implying that the parallax, which is used in \citet{bailerejones2021AJ....161..147B}, is unreliable. To conclude, given the low metallicity and considering our distance- and orbital estimates of the \jstar, it seems likely that it has an origin in the Galactic halo or thick disc.

\section{Conclusions}
In this work we present an r-process enriched bulge star, 2MASS J18082459-2548444 (the \fstar). We show that it is high in both molybdenum with [Mo/Fe] = 0.62\,dex and europium with [Eu/Fe] = 0.78\,dex, whilst having otherwise overall bulge-like abundances for other heavy trans-iron elements. Furthermore, the low [Ce,La/Eu] ratio suggests that the heavier trans-iron elements in the \fstar\ were predominantly produced by the r-process, rather than the s-process. 

We compare the star to the previously published r-enriched star in the bulge, MASS J18174532–3353235 (the \jstar), and estimate both stars' distances and orbits. We find that the \fstar\ likely is confined and has an origin in the bulge, whilst the \jstar\ is more probable to have a halo or thick disc origin. Nonetheless, we can not firmly exclude that it resides in the bulge or that it has bulge origin, since distance measurements of stars at these distances in general come with large uncertainties.

It would be beneficial to measure the ruthenium (Ru,\,Z\,=\,44) abundance of the \fstar, which has a 7\,\% p-process and 59\,\% contribution from the r-process at solar metallicities. The \jstar\ is extremely high in Ru, [Ru/Fe]\,=\,1.58\,dex, which could suggests that intermediate mass elements such as Ru and Mo are produced in a distinct way (possibly the p-process in the case for Ru and Mo). However, all Ru-lines in the optical, such as the 5309 Å and 5699\,Å line used in \citet{johnson2013ApJ...775L..27J}, lie outside of the spectrum range of 5800-6800\,Å for the \fstar, and new high-resolution spectra covering the whole optical region would be beneficial to be able to determine more trans-iron elements. It is key to use high resolution, high S/N optical spectra in the search for these r-process enhanced stars; the r-process signatures are difficult to infer in the infrared \citep[so far, only the r-process element ytterbium has had successful abundance determination in the infrared from high-resolution IGRINS spectra, see e.g.][]{montelius22}.

The discovery of one r-process enhanced star that is a member of the bulge does not by itself provide insight into the bulge formation process. What it does is to demonstrate, despite the significantly higher metallicities and implied greater rate of enrichment, that it was possible for sufficient r-process elements to be produced that a star like the \fstar\ could be discovered in the inner bulge. It is interesting to ask whether the tail of stars toward [Fe/H] $\sim -1$\,dex reflects an early period of enrichment in the history of the inner bulge/bar, and as such, whether other signatures of stochastic production of elements in the early bulge might be found with additional studies. 

We do have evidence that the inner halo/bulge at $b=-11^\circ$ \citep{koch2006} is interesting in this regard and that even small surveys find stars with unusual composition, including s- and p-process enhancements \citep{koch2016,koch2019}. However, it is important to emphasise that these stars, along with other recent surveys that emphasise the discovery of stars with [Fe/H]$<-2$\,dex \citep[e.g. EMBLA and Pristine Inner Galaxy Survey,][respectively]{howes2016,arentsen2020} explore a different potential discovery space, that of the early inner halo. The importance of the \fstar\ is the strong evidence for its presence in the inner bulge, and its [Fe/H]$>-1$\,dex. The discovery of other stars similar to it, and residing in the tail toward $-1$\,dex of the primary bulge metallicity distribution, might preserve the history of the enrichment event that gave rise to the chemical fingerprint associated with the formation of the bar. Alternatively, the low metallicity tail might preserve the record of an early merger. The discovery of additional similar stars and the detailed exploration of this population might give us new insight into the early formation history of the bulge. 


\begin{acknowledgements}
We thank the anonymous referees for comments and suggestions that helped improve the manuscript. We thank Paul McMillan for discussions on Galactic potentials and distance estimates. R.F's research is supported by the Göran Gustafsson Foundation for Research in Natural Sciences and Medicine. R.F.\ and N.R. acknowledge support from the Royal Physiographic Society in Lund through the Stiftelse Walter Gyllenbergs fond and Märta och Erik Holmbergs donation. N.N and M.S acknowledges the funding of the BQR Lagrange. R.M.R.\ acknowledges the financial support and hospitality of the Observatoire de Cote d'Azur. B.T.\ acknowledges the financial support from the Japan Society for the Promotion of Science as a JSPS International Research Fellow. This work has made use of data from the European Southern Observatory (ESO) and the European Space Agency's (ESA) {\it Gaia} mission (\url{https://www.cosmos.esa.int/gaia}), processed by the {\it Gaia} Data Processing and Analysis Consortium (DPAC,
\url{https://www.cosmos.esa.int/web/gaia/dpac/consortium}). Funding for the DPAC as been provided by national institutions, in particular the institutions participating in the {\it Gaia} Multilateral Agreement. 
\textit{Software:} \texttt{NumPy} \citep{numpy}, \texttt{Matplotlib} \citep{matplotlib2007CSE.....9...90H}, \texttt{GalPy} \citep[\url{http://github.com/jobovy/galpy},][]{galpybovy2015ApJS..216...29B}, \texttt{AstroPy} \citep{astropy2013A&A...558A..33A}.
\end{acknowledgements}

\bibliography{references}{} 
\bibliographystyle{aa} 

%
%

\end{document}